\documentclass[preprint,12pt]{elsarticle}
\usepackage{latexsym}
\usepackage{amssymb}
\usepackage{fullpage}
\usepackage{graphicx}
\usepackage[applemac]{inputenc}
\usepackage{float}
\usepackage{fancyhdr}
\usepackage{amsfonts}
\usepackage{amsmath}
\usepackage{color}
\usepackage{mathrsfs}
\usepackage{bm}
\usepackage{epsfig}

\usepackage{multirow}
\def\beq{\begin{equation}}
\def\eeq{\end{equation}}
\def\beqa{\begin{eqnarray}}
\def\eeqa{\end{eqnarray}}
\newcommand{\lsim}{\mbox{\raisebox{-.6ex}{~$\stackrel{<}{\sim}$~}}}

\def\paras{\vskip 0.2cm}

\def\parab{\vskip 0.8cm\noindent}

\def\cm{\,{\rm cm}}
\def\m{\,{\rm m}}
\def\s{\,{\rm s}}

\def\pc{\,{\rm pc}}
\def\kpc{\,{\rm kpc}}
\def\erg{\,{\rm erg}}

\def\tev{\,{\rm TeV}}
\def\mhz{\,\rm MHz}
\def\half{\frac{1}{2}}

\def\exp{{\rm exp}}

\def\fourthird{\frac{4}{3}}

\def\rem{r_{\rm em}}
\def\remCrab{r_{\rm em, Crab}}

\def\gray{$\gamma$-ray\ }

\def\mgro{MGRO J2019+37}

\def\me{m_e}
\def\mecsq{\me c^2}

\newcommand\mathcalLnuIC{{\mathcal L}_\nu^{\rm IC}}

\newcommand\mathcalLnuSy{{\mathcal L}_\nu^{\rm Sy}}

\newcommand\Fnusync{F_\nu^{\rm sy}}

\newcommand\FnuICCMB{F_\nu^{\rm IC,CMB}}
\newcommand\FnuSSC{F_\nu^{\rm SSC}}
\newcommand\FGMRT{F_{610\mhz}^{\rm GMRT}}

\newcommand\FMGRO{F_{10\tev}^{\rm MGRO}}
\newcommand\FCrabTeV{F_{10\tev}^{\rm Crab}}
\def\dnedgamma{\frac{dn_e}{d\gamma}}

\newcommand\gsync{g_{\rm sy}}

\newcommand\gICCMB{g_{\rm IC,CMB}}
\newcommand\gSSC{g_{\rm SSC}}
\newcommand\nuFnuCrab{\nu F_\nu^{\rm Crab}}
\newcommand\nuFnuMGRO{\nu F_\nu^{\rm MGRO}}
\newcommand\BCrab{B_{\rm Crab}}
\newcommand\Ee{E_e\,}
\newcommand\EeCrab{E_{e,{\rm Crab}}}
\newcommand\mathcalF{\mathcal F}
\newcommand\mathcalP{\mathcal P}
\newcommand\mathcalPset{\{\mathcalP\}}
\newcommand\xiCrab{\xi_{\rm Crab}}
\newcommand\xiMGRO{\xi_{\rm MGRO}}
\def\Ae{A_e}
\def\gammamax{\gamma_{\rm max}}

\begin{document}
\begin{frontmatter}

\title{Constraints on the synchrotron self-Compton mechanism of TeV 
gamma ray emission from the Milagro TeV source MGRO J2019+37 within the 
pulsar wind nebula scenario} 

\author{Lab Saha$^{1}$, Pijushpani Bhattacharjee$^{2}$}

\address{$^1$Tata Institute of Fundamental Research, Homi Bhabha Road, 
Colaba, Mumbai-400005, INDIA\\
$^2$ Saha Institute of Nuclear Physics, 1/AF Bidhannagar, 
Kolkata-700064, INDIA\\}
 
%
%
\begin{abstract}
Origin of the TeV gamma ray emission from MGRO J2019+37 
discovered by the Milagro experiment is investigated within the 
pulsar wind nebula (PWN) scenario using multiwavelength 
information on sources suggested to be associated with this object. We 
find that the synchrotron self-Compton (SSC) mechanism of origin of 
the observed TeV gamma rays within the PWN scenario is severely 
constrained by the upper limit on the 
radio flux from the region around MGRO J2019+37 given by the Giant 
Metrewave Radio Telescope (GMRT) as well as by the x-ray flux upper 
limit from SWIFT/XRT. Specifically, for the SSC mechanism to explain the 
observed TeV flux from \mgro\ without violating the GMRT 
and/or Swift/XRT flux upper limits in the radio and x-ray regions, 
respectively, the emission region must be extremely 
compact with the characteristic size of the emission region restricted 
to $\lsim{\mathcal O}(10^{-4}\pc)$ for an assumed distance
of $\sim$ few kpc to the source. This is at least four   
orders of magnitude less than the characteristic size of the emission 
region typically invoked in explaining the TeV emission through 
the SSC mechanism within the PWN scenario. On the 
other hand, inverse Compton (IC) scattering 
of the nebular high energy electrons on the cosmic microwave background 
(CMB) photons can, for reasonable ranges of values of various 
parameters, explain the observed TeV flux without violating the GMRT 
and/or SWIFT/XRT flux bounds. 

\end{abstract}
\end{frontmatter}

\section{Introduction}
The Cygnus region of the Galaxy hosts a number of extended, 
unidentified sources of TeV $\gamma$-ray emission, the most prominent of 
which is \mgro\  discovered by the Milagro 
experiment~\cite{2007ApJ...658L..33A,2007ApJ...664L..91A}. Detailed 
analysis~\cite{2012ApJ...753..159A} of the observational data on
this object collected during the period 2005--2008 gives a detection of 
this source with a statistical significance in excess of $12\sigma$ 
between 1 and 100 TeV. The measured 
flux~\cite{2012ApJ...753..159A} from this source is 
$7^{+5}_{-2}\times10^{-10}\s^{-1}\m^{-2}\tev^{-1}$ (with a $\sim30\%$ 
systematic uncertainty) at 10 TeV with a spectrum that is best described 
by a power-law with a spectral index of $2.0^{+0.5}_{-1.0}$ (with a 
systematic uncertainty of $\sim 0.1$) and an exponential cutoff at an 
energy $E_c=29^{+50}_{-16}\tev$. 

\paras 
Although no confirmed counterparts of the TeV source \mgro\ at lower 
energies are known, several possible associations with 
other observed sources have been suggested. The emission from \mgro\ may 
be due to either a single extended source or several unresolved sources. 
The EGRET sources 3EG J2021+3716 and 3EG J2016+3657 are positionally 
close to \mgro, and thus could be the GeV counterparts of \mgro\  
if it is a multiple source. At the same time, the EGRET source 3EG 
J2021+3716 is suggested to be associated with the radio and GeV pulsar 
PSR J2021+3651 (and its associated pulsar wind nebula PWN G75.2+0.1 
\cite{2002ApJ...577L..19R,2004ApJ...612..389H}) observed at GeV 
energies by AGILE (AGL J2020.5+3653)~\cite{2008ApJ...688L..33H} 
as well as FERMI (2FGL J2021.0+3651)~\cite{2009ApJS..183...46A}.  
A SWIFT/XRT observation~\cite{2007ATel.1097....1L} was also done 
within the positional uncertainty region of \mgro\ reported in       
\cite{2007ApJ...658L..33A,2007ApJ...664L..91A} and three x-ray sources 
were reported in the region with a total x-ray flux corresponding to 
$\nu F_\nu \sim 8.1\times10^{-14}\tev\,\cm^{-2}\,\s^{-1}$ in the 2--10 
keV energy region, which can, therefore, be taken as an upper limit on 
the possible x-ray flux from any x-ray counterpart of \mgro\  in this 
energy region. In addition, a wide-field deep radio survey of the \mgro\ 
region at 610 MHz was made by the Giant Metrewave Radio Telescope 
(GMRT)~\cite{2009A&A...507..241P}, 
yielding no detectable radio source, thus giving a conservative 
upper limit of $\sim$ 1.0 mJy on the radio flux from any point-like 
radio counterpart of \mgro. Recently, a deep very high energy (VHE)  
observation by the VERITAS experiment \cite{0004-637X-788-1-78} has 
resolved the VHE emission from the region of MGRO J2019+37 into two VHE 
sources. One of these, namely, VER J2019+378, which is also positionally 
coincident with the energetic GeV pulsar PSR J2021+3651 and its PWN, 
has been suggested to be associated with the bulk of the TeV emission 
from \mgro.  

\paras 
In this paper, we study the implications of a scenario in which 
the observed TeV \gray emission from \mgro\ arises from a Pulsar Wind 
Nebula (PWN) type source. Pulsar Wind Nebulae (PWNe) (see, e.g., 
\cite{2006ARA&A..44...17G} for a review), a well-known example of 
which is the Crab Nebula (see \cite{2008ARA&A..46..127H} for a review), 
are known to be sources of very high energy gamma rays extending 
to TeV energies (see, e.g., 
\cite{Aharonian_book_2004,2012ASPC..466..167K} for reviews). The TeV 
photons are thought to be emitted mainly through (a) inverse Compton 
(IC) interaction of high energy electrons with the low 
energy synchrotron photons emitted by the electrons themselves
in the ambient magnetic field in the nebula --- the so-called 
synchrotron self-Compton (SSC) mechanism, and/or (b) IC interaction with 
the photons constituting the cosmic microwave background (CMB) and 
infrared 
background --- hereafter referred to as the ``IC-CMB" mechanism. The 
high energy electrons themselves are thought to be 
accelerated in the wind termination shock where the relativistic 
wind from the pulsar residing within the nebula is stopped by the 
nebular material. In principle, in addition to electrons high energy 
protons (and in general heavier nuclei) may be also accelerated 
~\cite{1996MNRAS.278..525A,1997PhRvL..79.2616B,2003A&A...402..827A,2003A&A...405..689B}, 
which can produce high energy photons through decay of neutral pions 
produced in inelastic $p$-$p$ collisions. In this paper we 
shall restrict our attention to the leptonic scenario, i.e., we 
assume that the TeV emission is due to 
electrons. We use multiwavelength data and flux upper limits from 
observations in the region around \mgro\ including the radio upper limit 
given by GMRT~\cite{2009A&A...507..241P}, x-ray flux upper limit 
from SWIFT/XRT observations \cite{2007ATel.1097....1L},
GeV observations by FERMI~\cite{2009ApJS..183...46A}, 
EGRET~\cite{1999ApJS..123...79H} 
and AGILE~\cite{2008ApJ...688L..33H} and the TeV data from 
Milagro~\cite{2007ApJ...658L..33A,2007ApJ...664L..91A,2012ApJ...753..159A} 
and VERITAS~\cite{0004-637X-788-1-78} to study the implications of both 
the SSC and IC-CMB mechanisms for the macroscopic parameters of the 
underlying PWN, namely, its energetics (e.g., the total 
energy contained in the high energy electrons), the characteristic size 
of the emission region and the ambient magnetic field within the nebula. 

\paras
We find that the SSC mechanism of origin of the observed TeV gamma rays 
from \mgro\ is severely constrained by the GMRT upper limit on the
radio flux from the region around MGRO J2019+37 as well as by the 
SWIFT/XRT x-ray flux upper limit. Specifically, for the 
SSC mechanism to explain the observed TeV flux from \mgro\ without 
violating the GMRT and/or Swift/XRT flux upper limits, the emission 
region within the PWN must be extremely
compact with the characteristic size of the emission region restricted
to $\lsim{\mathcal O}(10^{-4}\pc)$ for an assumed distance
of $\sim$ few kpc to the source. This is at least four
orders of magnitude less than the characteristic size of the emission
region typically invoked in explaining the TeV emission through
the SSC mechanism within the PWN scenario. On the other hand, the IC-CMB 
mechanism can, for reasonable ranges of values of various
parameters, explain the observed TeV flux without violating the GMRT
and/or SWIFT/XRT flux bounds. 

\paras
The reason for the upper limit on the size of the emission region in the 
SSC scenario is not hard to understand: Ignoring for the 
moment the details of the energy spectrum of the electrons, let $n_e$ 
denote the number density of the electrons and $\rem$ the characteristic 
radius of the (assumed spherical) emission region in the source. Recall 
that in the SSC mechanism the TeV photons are produced through IC 
interaction of the nebular high energy electrons with the synchrotron 
photons produced by the electrons themselves in the magnetic field in 
the nebular region. Since the number density of the synchrotron photons 
scales as $n_e$, the number density of the TeV photons produced by  
the electrons through the SSC mechanism roughly scales as $n_e^2$. Thus, 
for a given distance to the source, the emerging total TeV flux from the 
source scales as $n_e^2\, \rem^3$. The requirement of producing the 
observed TeV flux of \mgro, therefore, fixes the product 
$n_e\,\rem^{3/2}$. On the other 
hand, the photon fluxes in the radio and x-ray regions due to 
synchrotron emission by the electrons scale with the
product $n_e\,\rem^3$. Therefore, with the 
product $n_e\,\rem^{3/2}$ fixed by the observed TeV flux of \mgro, an 
upper limit on the radio flux given by GMRT or the x-ray flux given by 
SWIFT/XRT directly yields an upper limit on $\rem$. 
These arguments are elaborated upon more quantitatively in the 
following sections within the context of a simple power-law form of the 
energy spectrum of the electrons within the nebula. 

\paras 
The rest of the paper is organized as follows: First, in 
Section~\ref{sec:Multiwavelength_spectra} we set up the formulae to 
calculate the multiwavelength photon spectra and compare 
the resulting theoretically calculated multiwavelength photon spectra 
with the observed multiwavelength data and constraints pertaining to 
\mgro, and discuss the implications --- in particular the constraints on 
the characteristic size of the emission region --- for the SSC mechanism 
of production of the observed TeV flux from \mgro. Finally, 
we summarize our results and conclude in 
Section~\ref{sec:summary_conclusions}. 

\section{Multiwavelength photon spectra and constraints}
\label{sec:Multiwavelength_spectra}
For simplicity we shall assume a simple power-law form 
(with a high energy cutoff) of the energy spectrum of the high energy 
electrons within the system, namely,  
\begin{equation}
\dnedgamma=A_e \gamma^{-\alpha} \exp\left(-\gamma/\gammamax\right)\,,
\label{eq:dnedgamma}
\end{equation}
where $n_e$ denotes the number density of the electrons, $\gamma$ is  
the Lorentz factor of the electron, and $\Ae$, $\alpha$ and $\gammamax$ 
are parameters of the model. 

\paras
In general, the electron spectrum may be more complicated than the 
single power-law form assumed above. However, from the discussions 
given below it will be clear that the qualitative natures of 
the interrelationships between and constraints on the macroscopic parameters  
of the system we derive below are quite general and are independent of 
the exact form of the electron spectrum.  

\paras
It is clear that, unlike in the cases of well-studied specific PWNe 
such as the Crab Nebula for which the existence of detailed 
multiwavelength data allows one to determine the parameters of the 
underlying model by performing detailed spectral fits to 
observational data (see, e.g., \cite{2010A&A...523A...2M} in the 
case of the Crab nebula), it is not practical or even meaningful to 
attempt to ``determine" the parameters appearing in equation 
(\ref{eq:dnedgamma}) in the case of \mgro\ because of lack of such 
multiwavelength observational data. Instead, we shall focus on the 
plausible ranges of values of the most relevant parameters of the system 
by requiring that the resulting multiwavelength 
photon spectra be such as to be able to explain the observed TeV flux 
from \mgro\ without violating the upper limits on the x-ray and radio 
fluxes from the region around the object. 

\paras
For a given set of the electron parameters $\mathcalPset\equiv 
\{\alpha,\, \gammamax\,\}$ in (\ref{eq:dnedgamma}), the total energy 
contained in the electrons, $E_e=\fourthird\pi\rem^3 \mecsq\int
\gamma \dnedgamma d\gamma$, can be expressed as 
\beq
\Ee=\fourthird\pi\rem^3\, 
\Ae\,\mecsq\,\mathcalF\bigl(\{\mathcalP\}\bigr)\,,
\label{eq:E_e}
\eeq 
where $\rem$ is the radius of the (assumed) spherical region within which 
the electrons are assumed to be distributed uniformly, and $\mathcalF$ 
is a calculable function of the set of parameters $\mathcalPset$.    

\paras
Now, the energy spectrum of synchrotron photons produced by an electron 
of energy $\gamma\mecsq$ with a pitch angle $\theta$ in a magnetic field 
$B$ can be written as \cite{1970RvMP...42..237B}  
\beq
\mathcalLnuSy\equiv \left(\frac{d\mathcal E}{d\nu dt}\right)_{\rm Sy}
=\frac{\sqrt{3} e^3 B \sin\theta}{\mecsq}\,\, \frac{\nu}{\nu_c} 
\int^{\infty}_{\nu/\nu_c} ~~ K_{5/3}(x) dx\,,
\label{eq:L_nu_sy}
\eeq
where $e$ is the electron charge, 
\beq
\nu_c=\frac{3e\gamma^2}{4\pi m_e c}B \sin\theta\,,
\label{eq:nu_c}
\eeq
is the characteristic frequency of the emitted synchrotron radiation, 
and $K_{5/3}(x)$ is the modified Bessel function of fractional order 
5/3. In our calculations described below, we shall average over the 
electron pitch angle and adopt a value of $\sin\theta=\sqrt{2/3}$ 
\cite{2010A&A...523A...2M}.  

\paras
For simplicity, we shall work within the framework of the so-called 
``constant B-field" scenario 
\cite{1998ApJ...503..744H,2010A&A...523A...2M} and 
assume the magnetic field to be constant within the nebular region. 

\paras
Thus, for a source (assumed to be in steady state and emitting 
radiation isotropically) at a distance $D$, 
the synchrotron radiation flux (energy/area/time/frequency) at 
frequency $\nu$ at earth can be written as 
\beqa
\Fnusync & = & \frac{1}{4\pi D^2}\, \fourthird\pi\rem^3\, \int_1^\infty 
\mathcalLnuSy \, \dnedgamma \, d\gamma \, \nonumber\\ 
 & = & \frac{1}{4\pi D^2}\, \frac{\Ee}{\mecsq}\, \frac{e^3B}{\mecsq}\, 
\gsync\bigl(\nu, B, \mathcalPset\bigr)\,,
\label{eq:Fnusync}
\eeqa 
where we have used equations (\ref{eq:L_nu_sy}), (\ref{eq:dnedgamma}) 
and (\ref{eq:E_e}) in the second line and written the equation 
in a suggestive form by extracting the explicit 
dependence on dimensionful quantities thus making 
$\gsync\bigl(\nu, B, \mathcalPset\bigr)$ a 
numerically calculable dimensionless function of the indicated 
parameters. (Note that the combination $e^3B$ has the dimension of 
energy squared.) 

\paras 
Similarly, the energy spectrum of photons produced by an electron of 
energy $\gamma\mecsq$ due to IC scattering off a 
background of soft photons is given by~\cite{1970RvMP...42..237B} 
\beqa
\mathcalLnuIC &\equiv & \left(\frac{d\mathcal E}{d\nu dt}\right)_{\rm IC} \nonumber \\
&=&\frac{3}{4}\frac{\sigma_T c}{\gamma^2}\, h^2 \nu  
\int_{h\nu/(4\gamma^2)}^{h\nu} ~~
d \epsilon \frac{n_b(\epsilon)}{\epsilon}  ~f_{\rm IC}(\epsilon, \nu, 
\gamma)\,,
\label{eq:L_nu_IC}
\eeqa
where $\sigma_T$ is the Thomson cross section, $h\nu$ is photon energy 
after scattering, $n_b(\epsilon)\, d\epsilon\,$ is the number density 
of the background soft photons between energy $\epsilon$ and $\epsilon + 
d\epsilon$, and   
\beqa
f_{\rm IC}(\epsilon, \nu, \gamma) & = & 2 q \ln q + (1+2q)(1-q) + \half\, \, \nonumber \\ 
& & \times \frac{\left[4 \epsilon \gamma q/\left(\mecsq\right)\right]^2} 
{1+4\epsilon \gamma q /\left(\mecsq\right)} (1-q)\,,
\nonumber
\eeqa
with 
\beq
q = \frac{h\nu}{4 \epsilon 
\gamma^2 \left[1-h\nu/\left(\gamma\mecsq\right)\right]}\,.
\nonumber
\eeq

For the synchrotron self-Compton (SSC) process the target photons for 
the IC interaction of the high energy electrons are the synchrotron 
photons produced by the electrons themselves. As already mentioned in 
the Introduction, since the number density of these synchrotron photons 
scales with the electron number density, the SSC luminosity from the 
source will scale with the square of the electron number density. Thus, 
the total SSC flux at earth can be written as 
\beqa
\FnuSSC & = & \frac{1}{4\pi D^2}\, \fourthird\pi\rem^3\, \int_1^\infty
\mathcalLnuIC\, \dnedgamma\, d\gamma\, \nonumber\\
 & = & \frac{1}{4\pi D^2}\, \left(\frac{\Ee}{\mecsq}\right)^2\, 
\frac{\sigma_T c h}{\rem^3}\,
\gSSC\bigl(\nu, B, \mathcalPset\bigr)\,,
\label{eq:FnuSSC}
\eeqa
where again we have used equations (\ref{eq:dnedgamma}), (\ref{eq:E_e})  
and (\ref{eq:L_nu_IC}) (with target photon 
density $n_b$ in equation (\ref{eq:L_nu_IC}) replaced by the synchrotron 
photon density which scales with $A_e$), and 
$\gSSC\bigl(\nu, B, \mathcalPset\bigr)$ is another numerically 
calculable dimensionless function of the indicated parameters. Note the 
explicit dependence of $\FnuSSC$ on $\rem$, the radius of the emission 
region. 

\paras
Finally, for IC interaction on the universal CMB photons, with 
$n_b(\epsilon)$ in equation (\ref{eq:L_nu_IC}) replaced by 
\beq
n_b^{\rm CMB}\left(\epsilon\right)d\epsilon=\frac{8\pi}{(hc)^3} 
\frac{\epsilon^2}{\exp[\epsilon/k_BT]-1}d\epsilon\,,
\label{eq:n_b_CMB}
\eeq
with $T=2.725 {}^{\circ}$K, the IC-CMB flux at earth can be written as 
\beq
\FnuICCMB = \frac{1}{4\pi D^2}\, \frac{\Ee}{\mecsq}\,
\frac{\sigma_T h \nu^3}{c^2}
\gICCMB\bigl(\nu, \mathcalPset\bigr)\,,
\label{eq:FnuICCMB}
\eeq
where $\gICCMB\bigl(\nu, \mathcalPset\bigr)$ is again a 
numerically calculable dimensionless function.  

\paras
Equations (\ref{eq:Fnusync}), (\ref{eq:FnuSSC}) and (\ref{eq:FnuICCMB}) 
allow us to generate the multiwavelength spectra for our model of the 
\mgro\ for any chosen set of values of the various parameters involved. 
For a given source distance $D$ and the set of electron parameters 
$\mathcalPset$, there are three macroscopic parameters that control 
the multiwavelength spectra of the system, 
namely, the total energy contained in the electrons ($E_e$), the 
magnetic field ($B$) and the radius of the emission region ($\rem$). 
For judicious choices of values of these parameters, one can 
obtain multiwavelength spectra for our model of \mgro\ that 
provide reasonably good fit to the observed TeV data without violating 
the GMRT and SWIFT/XRT constraints. We shall discuss such 
multiwavelength spectra below. However, even without considering the  
full numerically generated multiwavelength spectra, we can see that 
within the context of the SSC mechanism of explaining the observed TeV 
data of \mgro, the GMRT and SWIFT/XRT flux upper limits impose an upper 
limit on the radius of the emission region, $\rem$. This can be 
simply seen as follows: 
\subsection{Upper limit on $\rem$ in the SSC scenario}
\label{subsec:rem_upper_limit}  
For an appropriate choice of the values of the magnetic field $B$ and 
the parameter set $\mathcalPset$, requiring 
that we be able to explain the observed TeV flux from \mgro\ 
at some energy, say, 10 TeV, by the SSC flux (\ref{eq:FnuSSC}), we get 
\beq
\frac{\Ee}{\mecsq}=\Bigl(\FMGRO\Bigr)^{1/2} \Bigl(\frac{4\pi D^2 
\rem^3}{\sigma_T c h}\Bigr)^{1/2} g^{-1/2}_{\rm SSC}\bigl(h\nu=10\tev, B, 
\mathcalPset\bigr)\,,
\label{eq:Ee_FMGRO_rel}
\eeq  
where $\FMGRO$ is the observed flux from \mgro\ at 10 TeV. But, at the 
same time, we must ensure that, for the same values of $B$ and the 
parameter set $\mathcalPset$, the synchrotron flux given by equation 
(\ref{eq:Fnusync}) at $\nu=610\mhz$ not exceed the radio flux upper 
limit given by GMRT~\cite{2009A&A...507..241P}, $\FGMRT$, from the 
region around the observed position of \mgro. This gives the condition 
\beq
\frac{1}{4\pi D^2}\, \frac{\Ee}{\mecsq}\, \frac{e^3B}{\mecsq}\,
\gsync\bigl(\nu=610\mhz, B, \mathcalPset\bigr) \leq \FGMRT\,, \nonumber
\eeq 
which, upon substituting for $\Ee$ from equation 
(\ref{eq:Ee_FMGRO_rel}), gives an upper limit on $\rem$:  
\beqa
\rem^{3/2} & \leq & \left(4\pi 
D^2\right)^{1/2}\, 
\bigl(\frac{\mecsq}{e^3B}\bigr)\, \left(\sigma_T c h\right)^{1/2}\, 
\FGMRT \,\, \Bigl(\FMGRO\Bigr)^{-1/2} \nonumber \\ 
& &    
g^{-1}_{\rm sy}\bigl(\nu=610\mhz, B, \mathcalPset\bigr)\,  
g^{1/2}_{\rm SSC}\bigl(h\nu=10\tev, B, \mathcalPset\bigr)\,. 
\,\,\,\,\,\,({\rm GMRT})   
\label{eq:rem_constraint_GMRT}
\eeqa

\paras
Similarly, one can derive an upper limit on $\rem$ from the SWIFT/XRT 
x-ray flux upper limit. The actual value of the upper limit on $\rem$ 
will be the lower of the two upper limits.  

\paras
Clearly, the upper limit on $\rem$ depends on the electron parameters, 
the magnetic field and on the distance to the source. Below we shall 
present the results of our detailed numerical calculations for 
various values of the parameters involved. However, 
rough estimates of the upper limit on $\rem$ can be 
obtained simply by comparing the observed TeV flux of \mgro\ and the 
GMRT (SWIFT/XRT) upper limit on its radio (x-ray) flux with the 
observed TeV and radio (x-ray) fluxes, respectively, of a known PWN 
system such as the Crab nebula, for example. Let us define the ratio 
\beq
\xi\equiv \frac{\nu\FnuSSC(h\nu=10\tev)}{\nu\Fnusync(\nu=610\mhz)}\,,
\label{eq:xi_def}
\eeq
where $\Fnusync$ and $\FnuSSC$ are given by equations (\ref{eq:Fnusync}) 
and (\ref{eq:FnuSSC}), respectively. Let us demand that the 
measured TeV flux of both Crab and \mgro\ be explained by the SSC 
process. Then, for a given set of the 
electrons' spectral parameters $\mathcalPset$ and magnetic field $B$, 
assumed same for the moment for both Crab and \mgro, it is easy to 
see, using equations (\ref{eq:xi_def}), 
(\ref{eq:Fnusync}), (\ref{eq:FnuSSC}) and (\ref{eq:Ee_FMGRO_rel}), that 
\beq
\left(\frac{\rem^{3/2}}{D}\right)_{\rm MGRO} = 
\left(\frac{\rem^{3/2}}{D}\right)_{\rm Crab}  
\left(\frac{\FCrabTeV}{\FMGRO}\right)^{-1/2} \frac{\xiCrab}{\xiMGRO}\,,
\label{eq:xiCrab-xiMGRO_reln}
\eeq
where the sub(super)scripts Crab and MGRO refer to quantities relevant 
to Crab and \mgro, respectively.  

\paras
From the observed multiwavelength spectral energy distribution of 
the Crab (see, e.g., \cite{2010A&A...523A...2M} for a compilation of 
the data; also see the Figures below) it can be seen that Crab 
emits comparable amount of energy at TeV and radio wavelengths, with 
$\nuFnuCrab(h\nu=10\tev)\simeq 1.5\times10^{-11}\,\erg\, \cm^{-2} 
\s^{-1}$ and 
$\nuFnuCrab(\nu=610\mhz)\simeq0.74\times10^{-11}\,\erg\, \cm^{-2} 
\s^{-1}$. In contrast, for \mgro, while the energy emitted at TeV 
energies is comparable with that for Crab, with $\nuFnuMGRO(h\nu=10\tev) 
\simeq 1.1\times10^{-11}\,\erg\, \cm^{-2} \s^{-1}$, the GMRT upper limit 
restricts the possible flux of \mgro\ in the radio region to 
$\nuFnuMGRO(\nu=610\mhz)\leq 6.1\times10^{-18}\,\erg\, \cm^{-2} 
\s^{-1}$, about 6 orders of magnitude less than the corresponding 
quantity for Crab at that frequency. Thus we have $\xiCrab\simeq 2$ and 
$\FCrabTeV/\FMGRO\simeq 1.36$, whereas $\xiMGRO\geq 1.8\times10^6$. 

\paras 
Using these numbers in equation (\ref{eq:xiCrab-xiMGRO_reln}) we get the 
constraint 
\beq
r_{\rm em, MGRO} \leq 9.68\times10^{-5}\,\, 
r_{\rm em, Crab}\,
\left(\frac{D_{\rm MGRO}}{D_{\rm 
Crab}}\right)^{2/3}\,.\,\,\,\,\,\,\,({\rm 
GMRT})
\label{eq:remMGRO_constraint_1} 
\eeq

\paras
The distance to \mgro\ is not precisely known. The radio and
GeV pulsar PSR J2021+3651 with its associated pulsar wind nebula PWN
G75.2+0.1
\cite{2002ApJ...577L..19R,2004ApJ...612..389H,2008ApJ...688L..33H,2009ApJS..183...46A}  
that has been suggested to be associated with \mgro\  is inferred to be 
at a distance of 3--4 kpc~\cite{2008ApJ...680.1417V}. Below, for our 
numerical calculations, we shall take 
$D_{\rm MGRO}=3\kpc$. For the Crab nebula we shall take 
$D_{\rm Crab}\sim2\kpc$ and $r_{\rm em, Crab}\sim 
1\pc$~\cite{2010A&A...523A...2M}. With these numbers, we get 
\beq
r_{\rm em, MGRO} \leq 1.3\times10^{-4}\pc\, 
\left(\frac{r_{\rm em, Crab}}{1\pc}\right) \left(\frac{D_{\rm 
MGRO}}{3\kpc}\right)^{2/3} 
\left(\frac{2\kpc}{D_{\rm Crab}}\right)^{2/3}\,. \,\,\,\,\,\,\,\,\,({\rm 
GMRT})
\label{eq:remMGRO_constraint_2}
\eeq
Similarly one can derive a rough estimate of the upper limit on $\rem$ 
using the SWIFT/XRT flux upper limit.  

\begin{figure*}[h!]
\centering
\begin{tabular}{c}
\includegraphics[width=0.72\textwidth,angle=270]{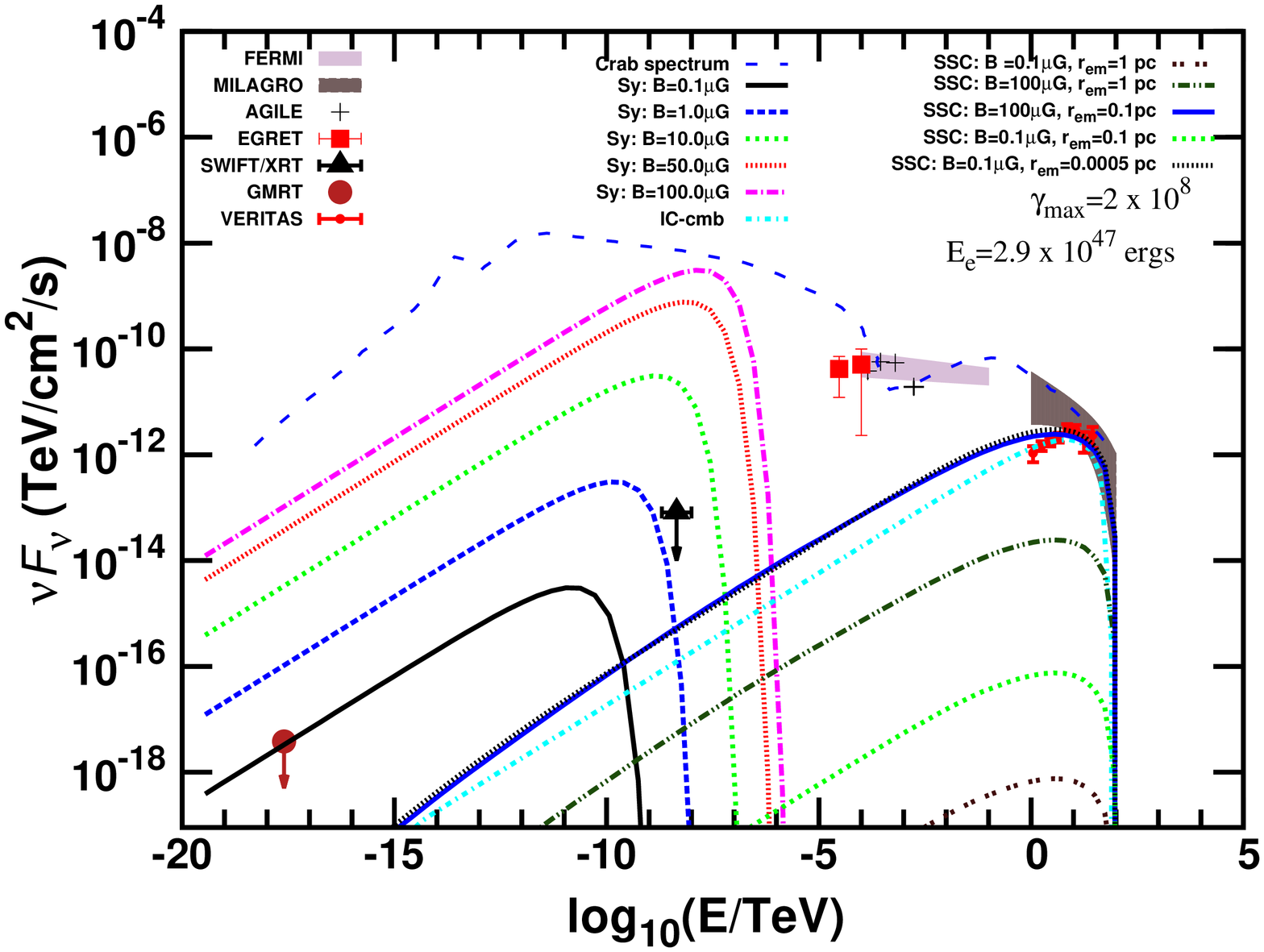}
\end{tabular}
\caption{The spectral energy distribution (SED) of our model of \mgro\  
from radio to TeV region in the PWN scenario. The curves spanning from 
radio to x-ray energies are the synchrotron spectra (marked by the 
key ``Sy") for various values of magnetic field ($B$) as shown. The 
curves going up to TeV energies and marked by the key ``SSC" are the 
synchrotron self-Compton spectra for selected values of 
the magnetic field and radius of the emission region ($\rem$) as shown. 
The curve marked "IC-CMB" is the spectrum produced through inverse 
Compton scattering of the electrons on the cosmic microwave background. 
The power-law index of the electron energy spectrum has been 
taken to be $\alpha=2$. The values of the exponential cutoff parameter 
of the electron spectrum ($\gammamax$) and the total energy contained in 
electrons ($\Ee$) are as specified. The distance to the source is taken 
to be $D=3\kpc$. Multiwavelength observational data and flux constraints 
from observations in the region around \mgro\  including the radio upper 
limit given by GMRT~\cite{2009A&A...507..241P}, x-ray flux upper 
limit from SWIFT/XRT observations~\cite{2007ATel.1097....1L},
GeV observations by FERMI~\cite{2009ApJS..183...46A}, 
EGRET~\cite{1999ApJS..123...79H} and 
AGILE~\cite{2008ApJ...688L..33H} and TeV observations by  
Milagro~\cite{2007ApJ...658L..33A,2007ApJ...664L..91A,2012ApJ...753..159A} 
and VERITAS~\cite{0004-637X-788-1-78} 
are shown. In addition, the SED of the Crab nebula 
(at a distance of $\sim$ 2 kpc) from 
radio to TeV energies \cite[taken from Ref.~][]{2010A&A...523A...2M} is  
also shown for comparison. 
}
\label{Fig:Fig1_B_0.1--100_rem_0.1-1-5e-4_Ee47}
\end{figure*}

\begin{figure*}[h!]
\centering
\begin{tabular}{c}
\includegraphics[width=0.72\textwidth,angle=270]{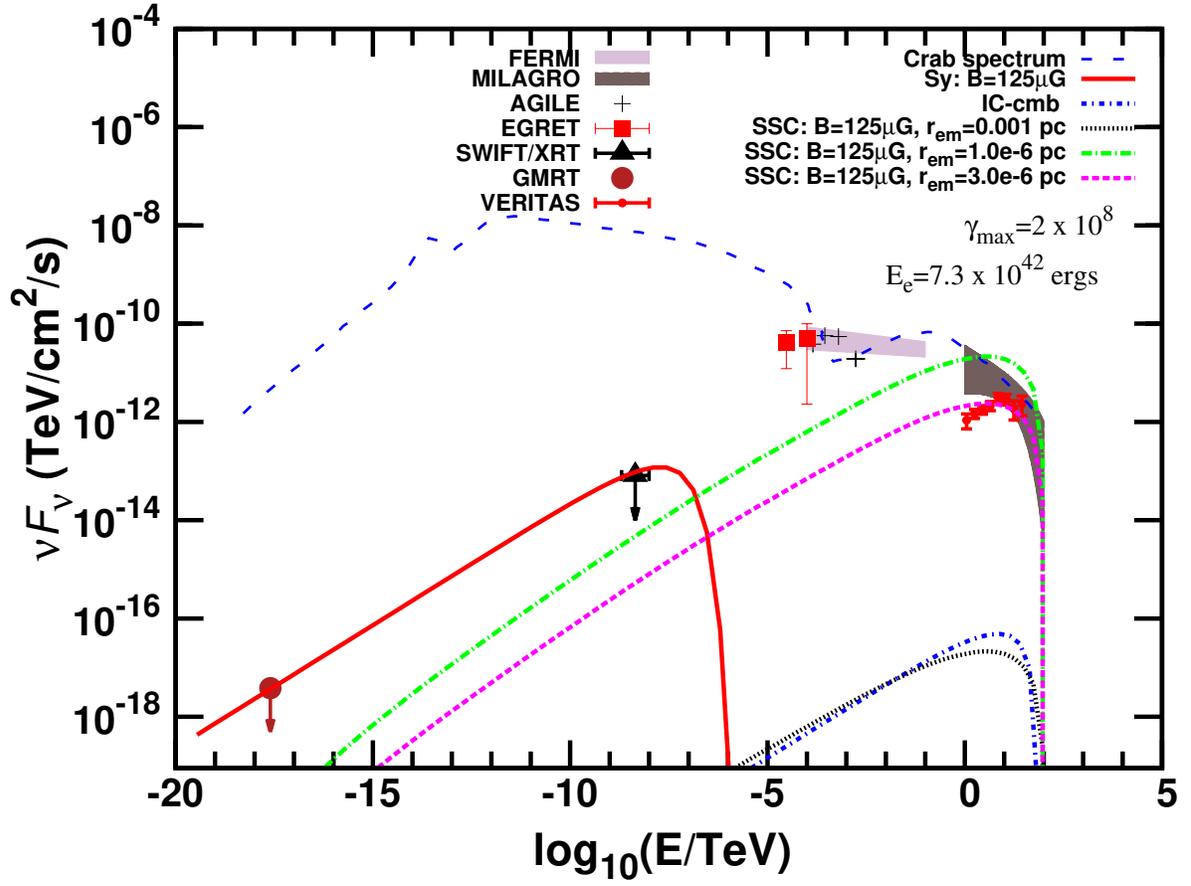}
\end{tabular}
\caption{Same as Figure \ref{Fig:Fig1_B_0.1--100_rem_0.1-1-5e-4_Ee47} 
but only for magnetic field $B=125\mu$G with $\Ee=7.3\times10^{42}\erg$   
and different values of $\rem$ as shown. 
}
\label{Fig:Fig2_B_125_rem_1e-3--e-6_Ee42}
\end{figure*}

\paras
These upper limit values indicate that if the spectral parameters of the 
electron population and the magnetic field of \mgro\ are same as  
those of the Crab nebula, then within the context of 
the SSC mechanism of production of TeV photons, the \mgro\  has to be a 
significantly more compact source than the Crab. Of course, 
the electron parameters and the magnetic field inside \mgro\ have no 
reason to be same as those in Crab. However, this does not alter the 
above general inference on the compactness of the source, the main 
reason for which is the strong constraint (upper limits) on the 
possible radio and x-ray fluxes 
from \mgro\  imposed by the GMRT and the SWIFT/XRT observations, 
respectively. To demonstrate this, we have generated a large set of 
multiwavelength spectra for our model of \mgro\ for a wide range of 
values of the parameters $\Ee$, $B$ and $\rem$ and electron 
parameters $\mathcalPset$ both for the SSC as well as IC-CMB mechanisms 
of production of TeV energy photons\footnote{We have included the 
contribution from IC scattering of the electrons with the interstellar 
radiation field (ISRF) (taken from Ref.~\cite{1983A&A...128..212M}) in 
addition to the CMB within the ``IC-CMB" contribution in our 
numerical calculations.}. Samples of such 
multiwavelength spectra are shown in Figures 
\ref{Fig:Fig1_B_0.1--100_rem_0.1-1-5e-4_Ee47} and 
\ref{Fig:Fig2_B_125_rem_1e-3--e-6_Ee42} for illustrating our main 
results. In these Figures the electron spectrum 
parameters have been taken to be $\alpha=2$ and $\gammamax=2\times10^8$ 
(corresponding to exponential cutoff energy of the electron spectrum at 
100 TeV); our basic results and conclusions do not change for other 
reasonable values of these parameters. 

\paras 
To set the scale for the possible 
ranges of values of the parameters $B$, $\Ee$ and $\rem$ for our 
numerical calculations, we note the typical values of these parameters 
invoked in 
explaining the multiwavelength emission from the Crab nebula, namely, 
$\EeCrab\simeq 5.3\times10^{48}\erg$, $\BCrab\simeq 125\mu$G and 
$\remCrab\simeq 1\pc$ (see, e.g., \cite{2010A&A...523A...2M}). We, 
however, keep in mind that the Crab is one of the most 
powerful PWNe, and the values of the above parameters for the \mgro\  
may be quite different from those for the Crab. Indeed, as 
we see from the Figures shown below, the values of these 
parameters required to explain the observed TeV data maintaining 
consistency with GMRT and SWIFT/XRT flux upper limits are rather 
different from those invoked for the Crab. 

\paras
From Figure \ref{Fig:Fig1_B_0.1--100_rem_0.1-1-5e-4_Ee47} we see that 
for a value of $\Ee=2.9\times10^{47}\erg$, which is about 5\% of 
$\EeCrab$ mentioned above, the GMRT and the SWIFT/XRT flux upper limits 
restrict the magnetic field within \mgro\ to below $0.1\mu$G, three 
orders of magnitude lower than $\BCrab$ mentioned 
above. For larger values of $\Ee$, the magnetic field has to be even 
lower. However, for $B=0.1\mu$G, we require a value of 
$\rem=5\times10^{-4}\pc$ (compared to $\remCrab\simeq 1\pc$) to explain 
the observed TeV data of \mgro\ with the SSC mechanism. The 
physical reason for this is that, for the chosen value of $\Ee$, with a 
value of $B$ as low as $0.1\mu$G the target synchrotron photon {\it 
density} is not large enough to produce the observed TeV flux through 
the SSC mechanism unless the size of the emission region is sufficiently 
small. Note that, for the same value of $\Ee$, a magnetic field of 
$100\mu$G can explain the TeV flux with a value of $\rem=0.1\pc$, but 
such a large value of $B$ produces synchrotron flux that violates the 
GMRT and SWIFT/XRT upper limits. Note also that, with the above value of 
$\Ee$ the IC-CMB mechanism can well explain the observed TeV flux. 

\paras
In general, for consistency with the GMRT and SWIFT/XRT flux upper 
limits, larger values of magnetic field require smaller values of the 
total energy contained in electrons, $\Ee$. For example, as shown in 
Figure \ref{Fig:Fig2_B_125_rem_1e-3--e-6_Ee42}, a magnetic field of 
$B=125\mu$G will produce synchrotron flux consistent with GMRT and 
SWIFT/XRT upper limits only for $\Ee\leq7.3\times 10^{42}\erg$, a value 
about six orders of magnitude lower than $\EeCrab$. However, for such a 
low value of $\Ee$, the target synchrotron photon density is again not 
large enough to produce the observed TeV flux through the SSC mechanism 
unless the emission region is sufficiently compact with $\rem\simeq 
3\times 10^{-6}\pc$. Note further that with such a low value of $\Ee$, 
the IC-CMB mechanism is unable to explain the observed TeV flux. 

\paras
In the above discussions we have made the simplifying assumption of the 
electrons --- and consequently the photons generated by them --- being 
uniformly distributed within a spherical ``emission region" of radius 
$\rem$ within the nebula. More realistically, the radius $\rem$ may be 
considered as a kind of characteristic length scale of a possible 
non-uniform spatial distribution of the electrons and/or the photons 
generated by them~\cite{2010A&A...523A...2M,1998ApJ...503..744H}. Also, 
the magnetic 
field inside the nebula is likely to be not spatially constant, but 
rather varying with a characteristic length scale similar to $\rem$. 
These details, however, are unlikely to change the general conclusion 
regarding the extreme compactness of the emission region of 
\mgro\  in the case of SSC mechanism of production of the 
observed TeV emission from \mgro\ within the context of the general PWN 
model of the source. 
\section{Summary and conclusions}
\label{sec:summary_conclusions}
To summarize, in this paper we have considered a PWN scenario of the 
origin of the observed TeV gamma ray emission from \mgro. We find that, 
while no lower energy counterparts of this object have yet been 
identified, the upper limits on possible radio and x-ray emissions from 
the source provided by the GMRT and SWIFT/XRT, respectively, 
already imply that the observed TeV emission of \mgro\  must originate 
from a highly compact region with characteristic size of the emission 
region $\lsim{\mathcal O}(10^{-4}\pc)$ if the observed TeV flux from 
\mgro\ is dominantly produced by the SSC mechanism. Such a compact size 
of the emission region is difficult to envisage within the usual PWN 
scenario. On the other hand, IC scattering of the high energy electrons 
on the CMB photons can  explain the observed TeV flux without 
violating the radio and x-ray flux upper limits given by GMRT and 
SWIFT/XRT, respectively, for reasonable ranges of values of 
various relevant parameters. 

\parab\paras
{\bf References}
\bibliographystyle{elsarticle-num}
\bibliography{mgro}

\vfill\eject

\end{document}